\documentclass[]{article}
\usepackage[utf8]{inputenc}
\usepackage{amsmath}
\usepackage[pdftex]{graphicx}
\usepackage{amscd}
\usepackage{calrsfs}
\usepackage{slashed}
\usepackage{amsfonts}
\usepackage{authblk}
\usepackage{float} 
\usepackage[section]{placeins}
\graphicspath{{figs/}}
\usepackage{caption}

\title{On Schwarzschild anti De Sitter and Reissner-Nördstrom wormholes}
\author[1]{Oscar Brauer \thanks{brauer@ciencias.unam.mx}}
\author[2]{Miguel Socolovsky \thanks{ socolovs@nucleares.unam.mx}}
\affil[1]{Facultad de Ciencias, Universidad Nacional Aut\'onoma de M\'exico}
\affil[2]{Instituto de Ciencias Nucleares, Universidad Nacional Aut\'onoma de M\'exico, Cd. Universitaria, 04510, Ciudad de M\'exico, M\'exico, and Max-Planck-Institut für Physik  (Werner-Heisenberg-
	Institut), Föringher Ring 6, 80805, München, Germany}

\providecommand{\keywords}[1]
{
	\small	
	\textbf{Keywords:} #1
}
\begin{document}
\date{}
\maketitle

\begin{abstract}
We discuss the wormholes associated with the four-dimensional Schwarzschild ($S_4$), Schwarzschild anti De Sitter ($SaDS_4$), and Reissner-Nördstrom ($RN_4$) black holes, in Schwarzschild, isotropic and Kruskal-Szekeres cordinates. The first two coordinate systems are valid outside the horizons, while the third one is used for the interiors. In Schwarzschild coordinates, embedding for $SaDS_4$ exists only for a finite interval of the radial  coordinate $r$, and similar restrictions exist for $RN_4$. The use of the K-S coordinates allows us to give an explicit proof of the  pinching-off of the bridges,  making them non-traversable. The case of the extreme Reissner-Nördstrom ($ERN_4$) is also discussed.
\end{abstract}

\keywords{Wormholes, Schwarzschild anti De Sitter, Reissner-Nordström, Black Holes}

\

PACS numbers: 04.70.-s, 04.70.Bw

\

\section{Introduction}
In the context of black hole theory, for each value of the Schwarzshild time $t\in (-\infty,+\infty)$ or each spacelike slice in a Kruskal-Szekeres (K-S) diagram with slope between -1 and +1 [1,2], there exists a hypersurface which connects the causally disconnected exterior regions, asymptotically flat in the cases e.g. of Schwarzschild ($S$) and Reissner-Nördstrom ($RN$) black holes ($BH'S$), or asymptotically anti De Sitter ($aDS$) or De Sitter ($DS$) in the cases of ($SaDS$) or ($SDS$) $BH's$. (We consider the universal cover of $aDS$ which makes it, and to $SaDS$, free of closed causal curves.) 
This hypersurface, which in the case of 4-dimensional spacetime and spherical symmetry we can imagine as a 2-dimensional surface by choosing the polar angle $\theta$ at the equatorial plane ($\theta=\pi/2$) is called wormhole [3] or Einstein-Rosen bridge [4]. Typically these bridges are non traversable, that is, no kind of particle (massive or massless) can pass through them from one exterior region to the other exterior region, because inside the future and past  horizons -where Schwarzschild coordinates do not hold and must be replaced by another set of coordinates, say K-S coordinates- a process of pinching-off of the wormhole occurs that forbiddes the passing of the particles and so it is responsible of the non traversability. For the $S$ wormhole this was first proved by Fuller and Wheeler [5]; a qualitative description can be found in Carroll [6], and a detailed description of the embedding in $\mathbb{E}^3$ and time evolution was recently exhibited by Collas and Klein [7]. The appearance of wormholes in black holes led to the proposal of their existence outside this context, as solutions of the Einstein's equations in the presence of matter satisfying non standard energy conditions, and being traversable [8]. For a general introduction and developement of the subject see Visser [9]. A more recent review of these developements can be found in Lobo [10].

\

In the present paper we describe in detail, through the use of K-S coordinates, the pinching-off process and consequent non traversability of the wormholes associated with the 4-dimensional spherically symmetric Schwarzschild anti De Sitter ($SaDS_4$) (subsection 4.1), Reissner-Nördstrom ($RN_4$) (subsection 4.2), and extremal Reissner-Nördstrom ($ERN_4$) (subsection 4.3) black holes. Since both $SaDS_4$ and $RN_4$ minus $(IV\cup IV^\prime)$ (see Figs. 9.a,b) do not contain closed causal curves (the 2nd. spacetime is globally hyperbolic [11]), one expects the corresponding wormholes to be non traversable to avoid the possibility of time travel through them. Section 2 is devoted to the description of the wormholes in Schwarzschild coordinates, while section 3 and its subsections does the same in isotropic coordinates. In these two cases the coordinates only cover the exterior regions i.e. outside the event horizons. In section 5 we briefly comment on recent developements on eternal traversable wormholes.

\

\textit{Note}. We use metrics with signature $(+,-,-,-)$, and the natural system of units $G=c=1$.   
\section{Schwarzschild coordinates: embeddings in $\mathbb{E}^3$}
In this section we discuss the embeddings in $\mathbb{E}^3$ of the wormholes or Einstein-Rosen (ER) bridges associated with the spherical symmetric Schwarzschild ($S_4$),  Schwarzschild anti De Sitter ($SaDS_4$), and Reissner-Nördstrom ($RN_4$) black holes (BH's), in terms of Schwarschild coordinates $(t,r,\theta,\varphi)$. These coordinates cover the exterior regions i.e. outside the event horizons,
\begin{equation}
r_h=2M
\end{equation}
for $S_4$,  
\begin{equation}
r_h=(Ma^2)^{1/3}\left(  \left(  1+\sqrt{1+a^2/27M^2}\right) ^{1/3}+ \left(  1-\sqrt{1+a^2/27M^2} \right)  ^{1/3} \right) 
\end{equation}
for $SaDS_4$, and  
\begin{equation}
r_+=M(1+\sqrt{1-(Q/M)^2})
\end{equation}
for $RN_4$, where M is the mass of the BH,  
$a=\sqrt{{{3}\over{-\Lambda}}}$ is the curvature radius associated with the cosmological constant $\Lambda<0$, and $Q^2=p^2+q^2>M^2$ is the sum of the squares of electric and (hypotetical) magnetic charges. Clearly, $r_h\to 2M$ as $a\to+\infty$ i.e. as $\Lambda\to 0_-$, and $r_+\to2M$ as $Q^2\to 0_+$.
\subsection{$S_4$}
The metric is 
\begin{equation}
ds^2_{S_4}= \left(  1-2M/r \right)  dt^2-{{dr^2}\over{1-2M/r}}-r^2d\Omega^2_2, \ r>2M,
\end{equation}
with $d\Omega^2_2=d\theta^2+sin^2\theta d\varphi^2$. 

\

For constant $t=t_0$ and $\theta=\pi/2$, the remaining 2-dimensional metric is
\begin{equation}
ds^2_{S_4}|_{t_0,\pi/2}=-\left( {{dr^2}\over{1-2M/r}}+r^2d\varphi^2\right) .
\end{equation}
The Euclidean metric in $\mathbb{R}^3$ in cylindrical coordinates $(r,\varphi,z)$ is 
\begin{equation}
dl_E^2=dr^2+r^2\varphi^2+dz^2= \left(  \left(  {{dz}\over{dr}} \right) ^2+1 \right)  dr^2+r^2d\varphi^2,
\end{equation}
where the last equality defines a 2-dimensional surface $z(r)$. Identifying $-ds^2_{S_4}|_{t_0,\pi/2}$ with $dl_E^2$ we obtain $({{dz}\over{dr}})^2+1={{1}\over{1-2M/r}}$ which leads to
\begin{equation}
z=\pm\sqrt{2M}\int_0^r{{dr^\prime}\over{\sqrt{r^\prime-2M}}}=\pm 2\sqrt{2M}\sqrt{r-2M},
\end{equation}
i.e. 
\begin{equation}
r=r(z)=2M+{{z^2}\over{8M}}.
\end{equation}
So, $r(0)=2M$ (minimum value) and $r(z)=r(-z)$. With $\varphi\in[0,2\pi)$, the embedded surface is the paraboloid in Fig. 1 [12] which, as $z\to\pm\infty$ ($r\to+\infty$), approaches to the Euclidean 2-plane. Behind this result there is a tacit \textit{extension} of the Swarzschild metric originally valid only in region $I$ of the Kruskal diagram in Fig. 2 to region $IV$ of the same diagram.

\begin{figure}[h]
	\centering
	\includegraphics[width=.6\linewidth]{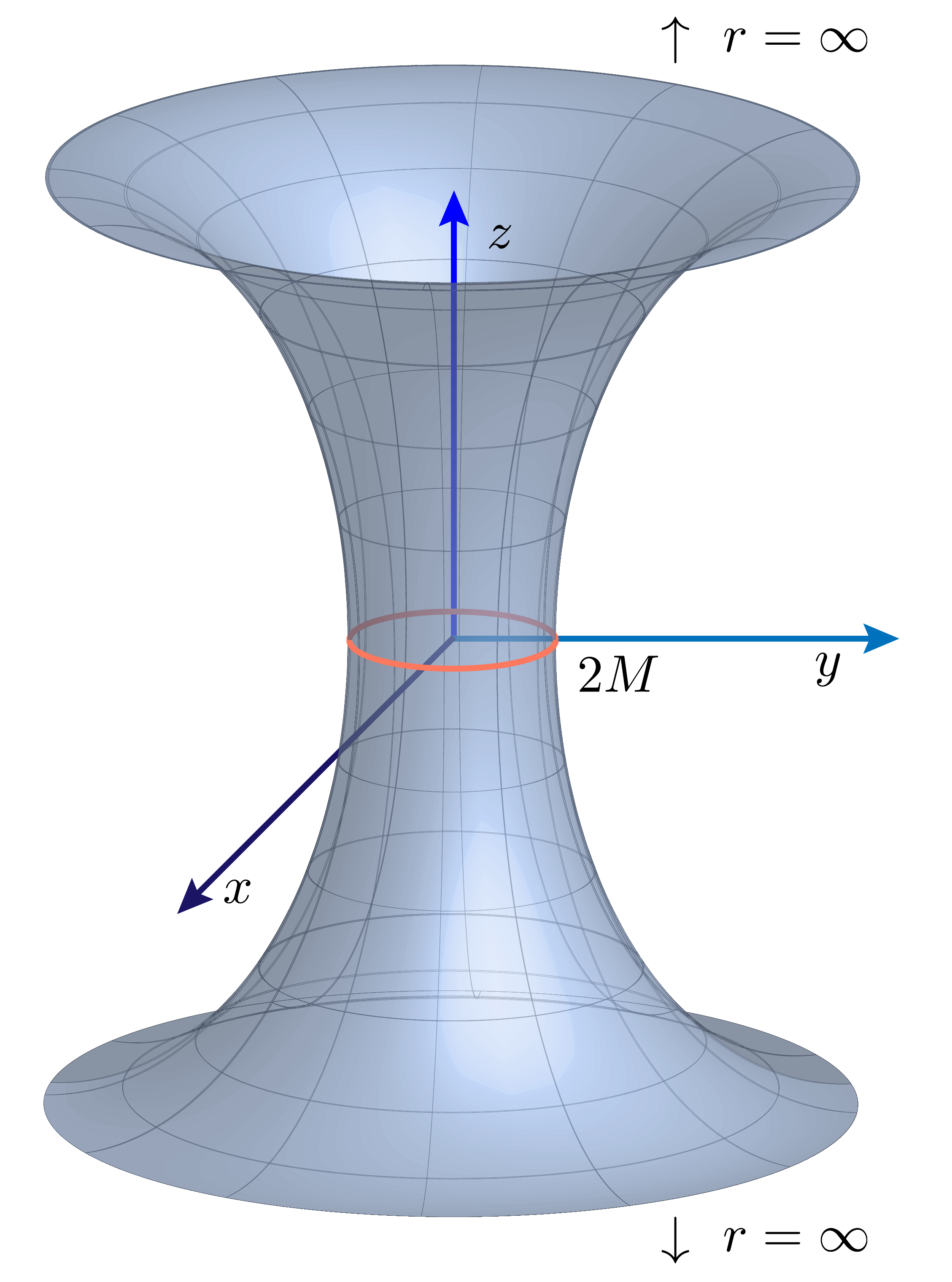}
	\caption{2-dimensional embedding of the $ S_{4} $ wormhole.}
\end{figure}

\begin{figure}[H]
	\centering
	\includegraphics[width=\linewidth]{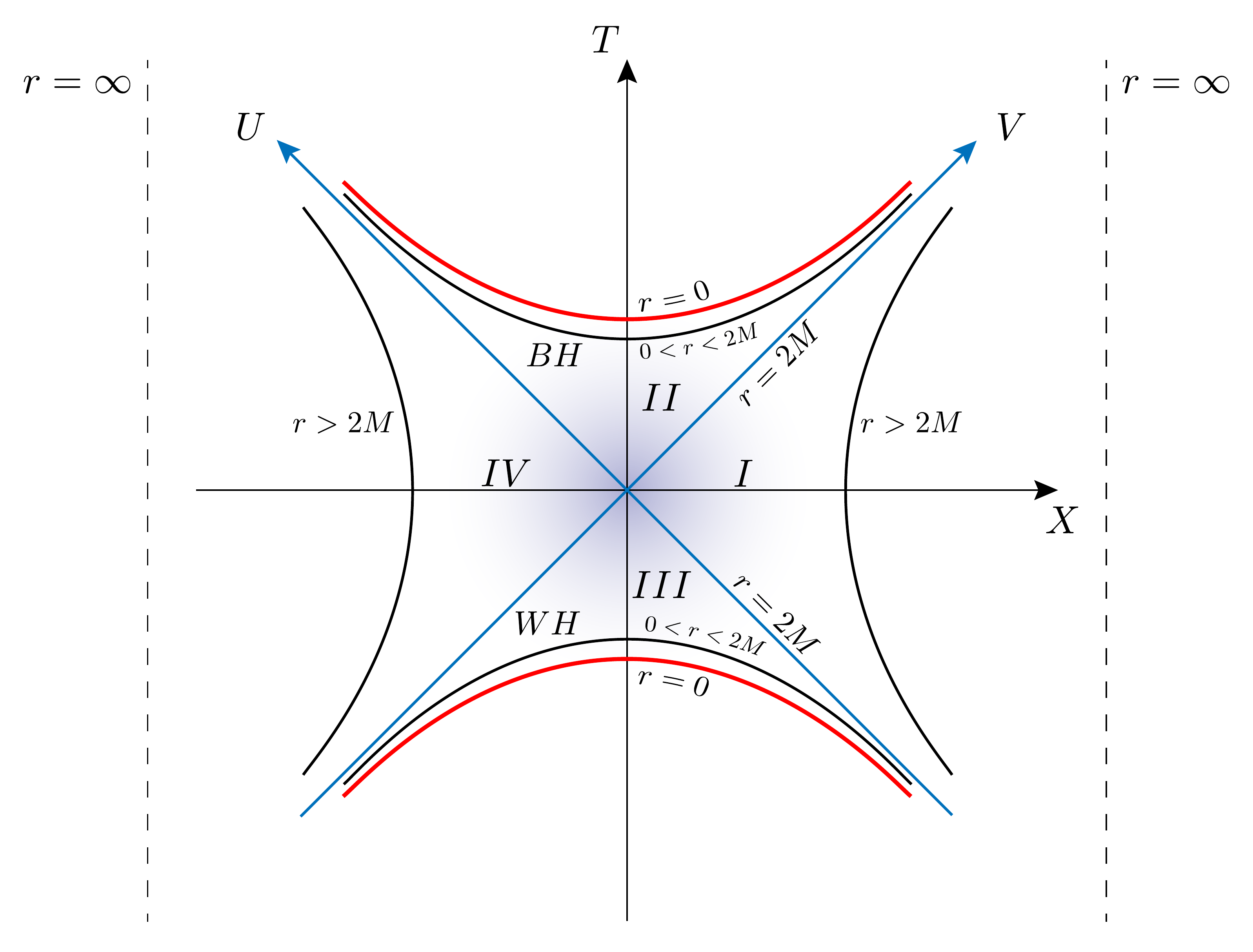}
	\caption{Kruskal-Szekeres diagram of $ S_{4} $.}
\end{figure}

\subsection{$SaDS_4$}
The metric is
\begin{equation}
ds^2_{SaDS_4}= \left(  1-{{2M}\over{r}}+{{r^2}\over{a^2}} \right)  dt^2-{{dr^2}\over{1-{{2M}\over{r}}+{{r^2}\over{a^2}}}}-r^2d\Omega^2_2.
\end{equation}
Again for constant $t=t_0$ and $\theta=\pi/2$, 
\begin{equation}
ds^2_{SaDS_4}|_{t_0,\pi/2}=- \left(  {{dr^2}\over{1-{{2M}\over{r}}+{{r^2}\over{a^2}}}}+r^2d\varphi^2 \right)  .
\end{equation}
Identifying $-ds^2_{SaDS_4}|_{t_0,\pi/2}$ with $dl_E^2$ in (5), we obtain
\begin{equation}
\left(  {{dz}\over{dr}} \right)  ^2={{1}\over{1-{{2M}\over{r}}+{{r^2}\over{a^2}}}}-1={{2Ma^2-r^3}\over{ra^2-2Ma^2+r^3}}.
\end{equation}
Since $ra^2-2Ma^2+r^3>0$ for $r>r_h$ and $({{dz}\over{dr}})^2>0$, then $2Ma^2-r^3>0$ which amounts to $r<(2Ma^2)^{1/3}$. I.e. the embedding of the surface in $\mathbb{E}^3$ only exists in the interval 
\begin{equation}
r_h<r< \left(  2Ma^2 \right)  ^{1/3}.
\end{equation}
(A straightforward calculation allows to prove that $r_h<(2Ma^2)^{1/3}$.) 

\

Using
\begin{equation}
2Ma^2=r_h(r_h^2+a^2),
\end{equation}
the embedding is given by the integral
\begin{equation}
z(r)=\int_{r_h}^r dr^\prime\sqrt{{{r_h(r_h^2+a^2)-  (r^\prime)^3}\over{(r^\prime)^3+a^2r^\prime-r_h(r_h^2+a^2)  
	}}}.
\end{equation}
\subsection{$RN_4$}
The metric is
\begin{equation}
ds^2_{RN_4}=fdt^2-{{dr^2}\over{f}}-r^2d\Omega^2_2
\end{equation}
with horizon function
\begin{equation}
f=f(r)=1-{{2M}\over{r}}+{{Q^2}\over{r^2}}.
\end{equation}
For $Q^2<M^2$, $f(r)$ has two roots:
\begin{equation}
r_\pm=M(1\pm\sqrt{1-Q^2/M^2})
\end{equation}
with $r_+$ the event horizon and $r_-$ a Cauchy horizon (see Fig. 3). For constant $t=t_0$ and $\theta=\pi/2$,
\begin{equation}
{ds^2_{RN_4}}|_{t_0,\pi/2}=-{{dr^2}\over{f}}-r^2d\varphi^2.
\end{equation}
Identifyng $-{ds^2_{RN_4}}|_{t_0,\pi/2}$ with $dl^2_E$ of eq. (6)
one obtains 
\begin{equation}
({{dz}\over{dr}})^2={{2M/r-Q^2/r^2}\over{1-2M/r+Q^2/r^2}}.
\end{equation}

\begin{figure}[H]
	\centering
	\includegraphics[width=.8\linewidth]{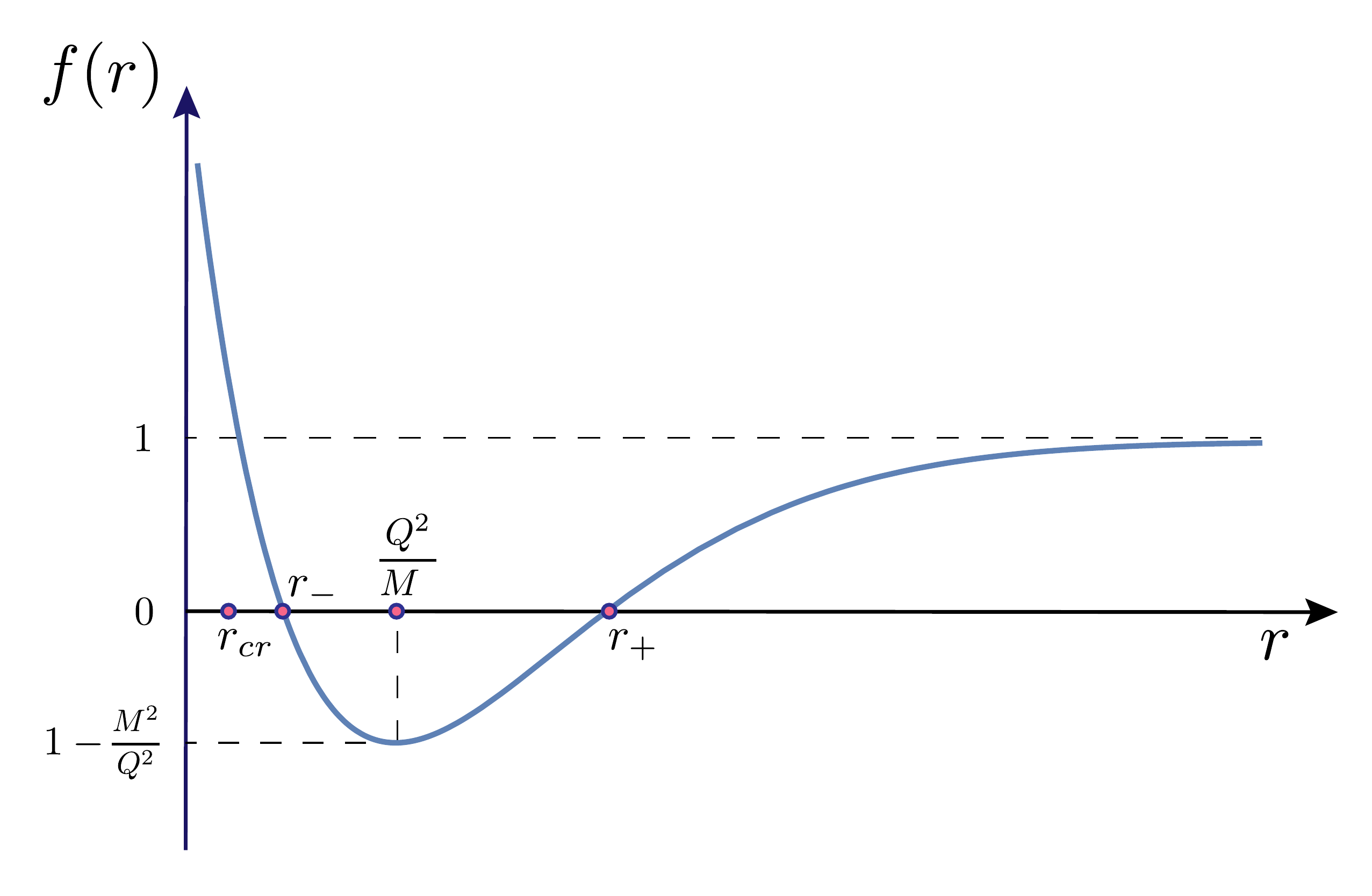}
	\caption{$ RN _{4}$ horizon function for $Q^{2}< M^{2}  $.}
\end{figure}

For the embedding outside the event horizon, $f(r)\geq 0$, then $2M/r-Q^2/r^2\geq 0$ which implies $r\geq Q^2/2M\equiv r_{cr}<r_+$; so the embedding in $\mathbb{E}^3$ exists for all $r\geq r_+$, and is given by the integral
\begin{equation}
z(r)=\pm\int_{r_+}^rdr^\prime\sqrt{{{2Mr^\prime-Q^2}\over{{r^\prime}^2-(2Mr^\prime-Q^2)}}}.
\end{equation}
Since $r_{cr}<r_-$, the embedding in $\mathbb{E}^3$ also exists in the range $r_{cr}\leq r\leq r_-$ and is given by the same integral with the lower limit replaced by $r_{cr}$ and an upper limit $r\leq r_-$. In the interval $r_-\leq r\leq r_+$, $f(r)\leq 0$, and so to keep $({{dz}\over{dr}})^2\geq 0$ it must be $r\leq r_{cr}$ which is outside the domain; so for $r_-\leq r\leq r_+$, in Schwarzschild coordinates, there is no embedding of the wormhole in $\mathbb{E}^3$. This region will be studied in Subsection 4.2 using Kruskal-Szekeres coordinates.
\subsection{$ERN_4$}
The extreme Reissner-Nördstrom case is defined by the condition $M^2=Q^2$ where both horizons coincide:
\begin{equation}
r_+=r_-=M.
\end{equation}
The metric becomes 
\begin{equation}
{ds^2_{ERN}}_4=(1-M/r)^2dt^2-{{dr^2}\over{(1-M/r)^2}}-r^2d\Omega_2^2
\end{equation}
(see Fig. 4). So
\begin{equation}
-{ds^2_{ERN}}_4|_{t_0,\pi/2}={{dr^2}\over{(1-M/r)^2}}+r^2d\varphi^2
\end{equation}
which identified with $dl_E^2$ gives
\begin{equation}
({{dz}\over{dr}})^2={{2Mr-M^2}\over{(r-M)^2}}.
\end{equation}
$({{dz}\over{dr}})^2\geq 0$ implies that the embedding of the wormhole exists only for $r\geq M/2$. As $r\to M$, $({{dz}\over{dr}})^2\to+\infty$ and therefore ${{dz}\over{dr}}\to\pm\infty$. 

\begin{figure}[H]
	\centering
	\includegraphics[width=.8\linewidth]{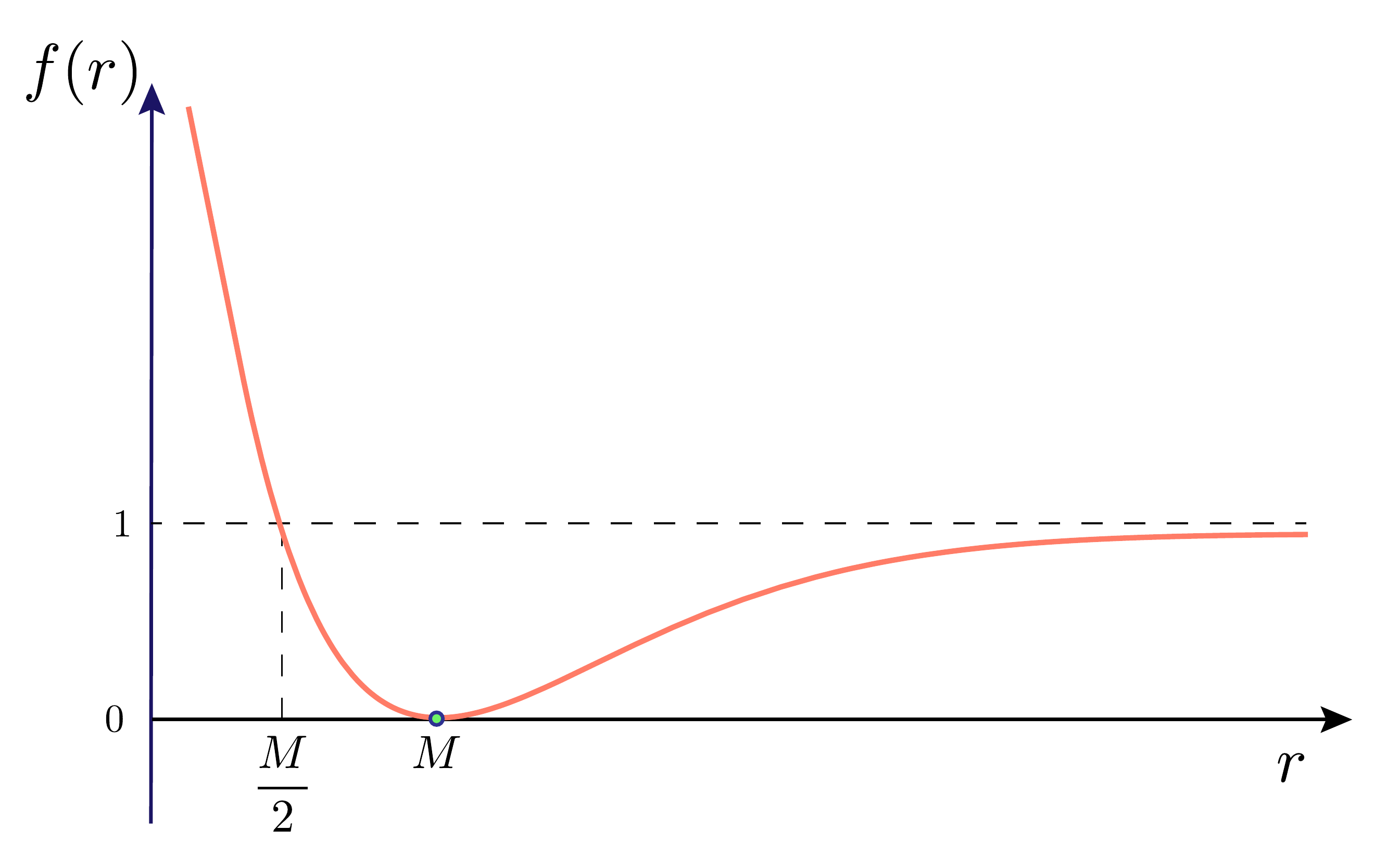}
	\caption{$ ERN_{4} $ horizon function.}
\end{figure}

The basic cell of the Penrose diagram of the $ERN_4$ spacetime (Fig. 5) only contains one asymptotically flat region ($I$) and the region $IV$ limited by the singularity at $r=0$ (we call it ``singular" region). Then the wormhole only lies in $I$ and $IV$. In other words, there is no parallel universe $I^\prime$ to which the universe $I$ is connected by the wormhole.

\begin{figure}[H]
	\centering
	\includegraphics[width=.5\linewidth]{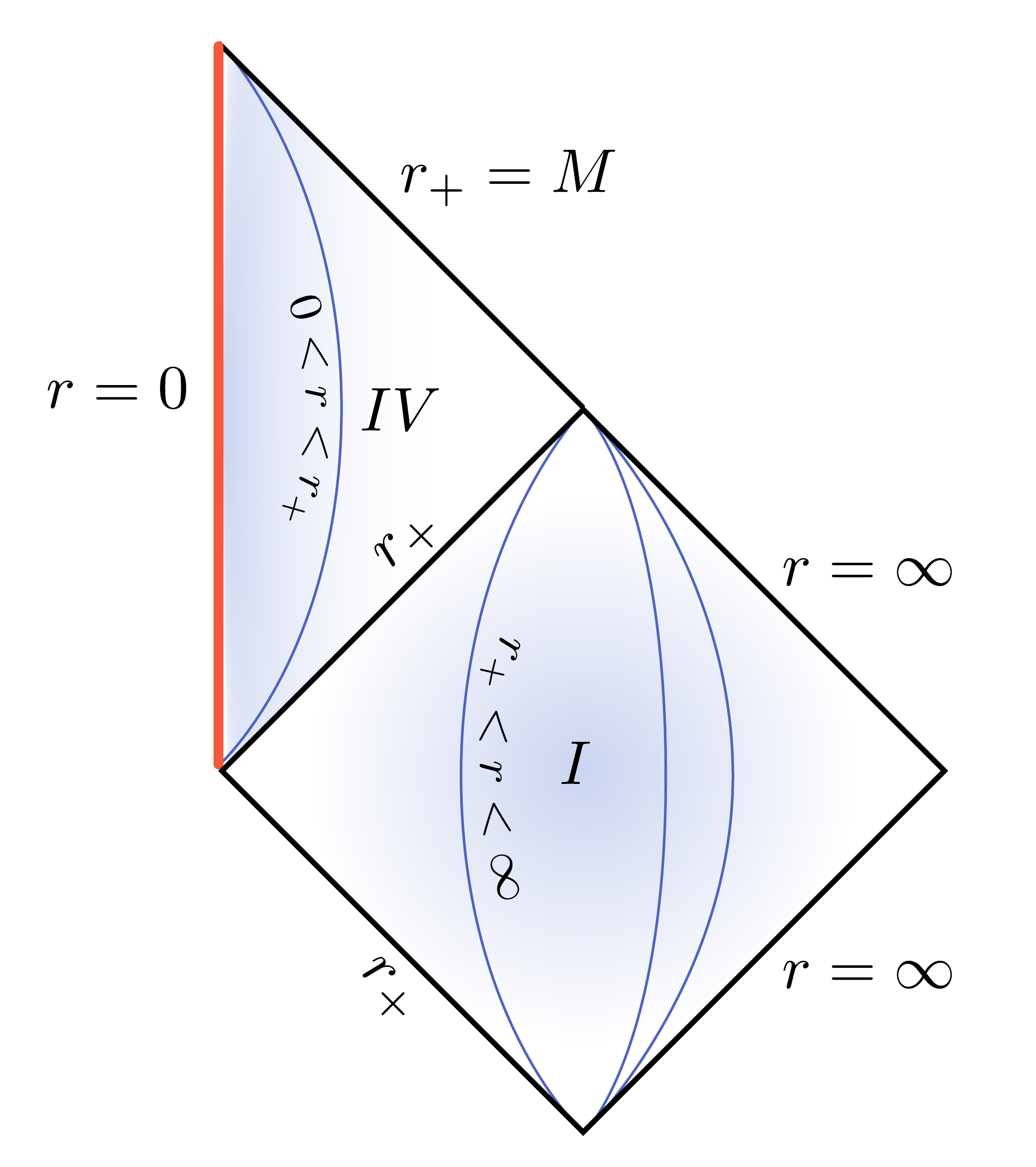}
	\caption{Basic cell of the Penrose diagram of $ ERN_{4} $.}
\end{figure}

\section{Isotropic coordinates}
Though our main interest is in the geometric description of the $SaDS_4$ and $RN_4$ wormholes, we first review the description of the $S_4$ wormhole in these coordinates. In all cases the analysis doesn't involve a standard embedding procedure; nevertheless, it gives a picture as if it were. Moreover, for the $SaDS_4$ ($RN_4$) case the picture is valid for all $r>r_h$ ($r>r_+$) i.e. for $\rho\in(0,+\infty)$, which makes a difference with respect to the embedding approach. 
\subsection{$S_4$} 
We define the coordinate $\rho$ through [13]
\begin{equation}
r= \left(  1+{{M}\over{2\rho}} \right)  ^2\rho
\end{equation}
with 
$\rho\in(0,+\infty)$, $[\rho]=[r]=[L]$. Then $r=\rho+{{M^2}\over{4\rho}}+M$ and so $r\to+\infty$ as $\rho\to 0_+$ and $\rho\to+\infty$. $r(\rho)$ has a minimum at $\rho={{M}\over{2}}$ with $r({{M}\over{2}})=2M$ (Fig. 6). Clearly then, the coordinates $(t,\rho,\theta,\varphi)$ only cover the exterior regions $(I)$ and $(IV)$ to the past and future horizons in the Kruskal diagram (Fig. 2). 

\begin{figure}[H]
	\centering
	\includegraphics[width=.7\linewidth]{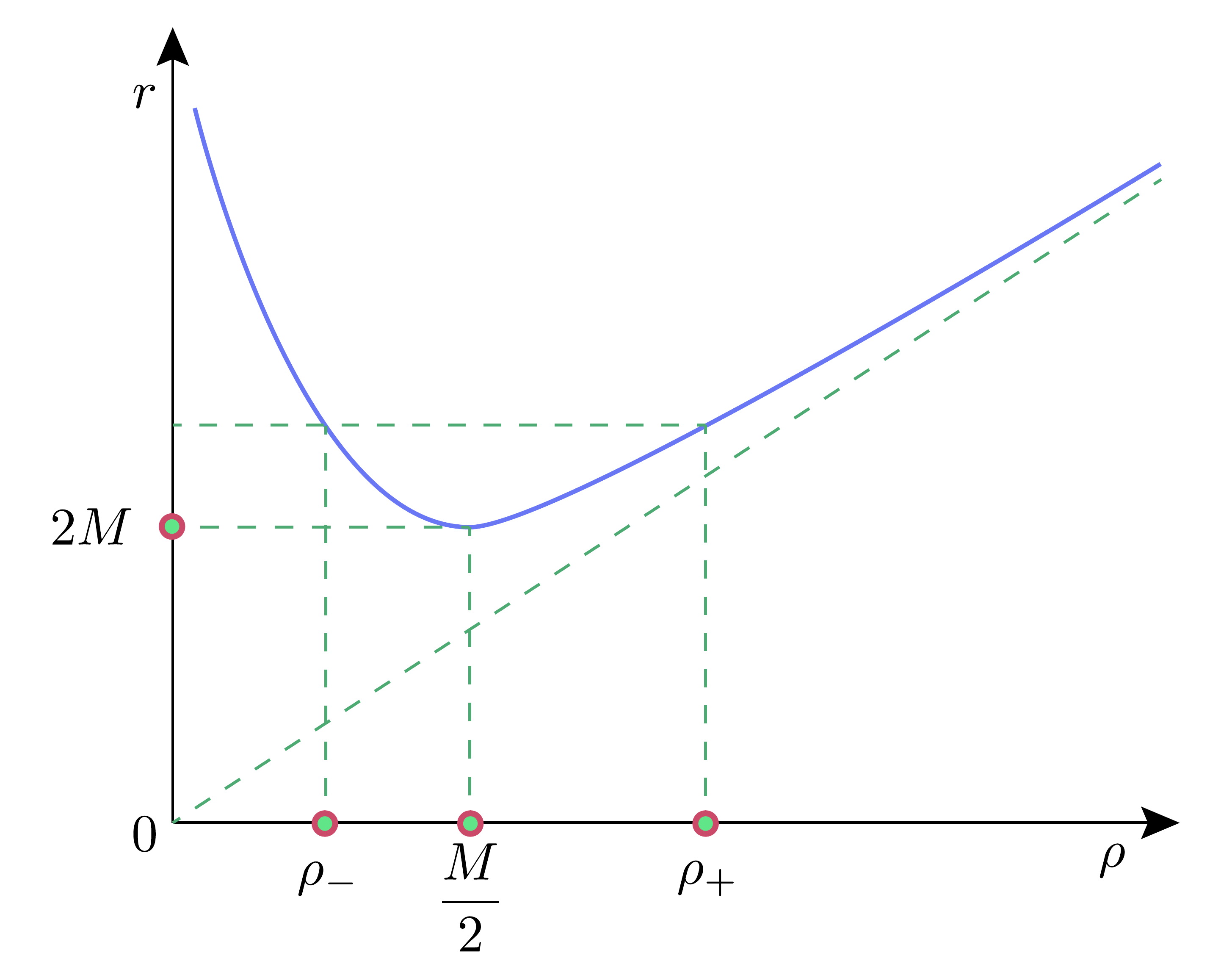}
	\caption{Relation between $ r $ and $ \rho $ for $ S_{4} $.}
\end{figure}

For each $r>2M$, there are two solutions of (25) given by
\begin{equation}
\rho_\pm={{r-M}\over{2}} \left(  1\pm\sqrt{1- \left(  {{m}\over{r-M}} \right)  ^2} \right)  , 
\end{equation}
with $\rho_\pm\to{{M}\over{2}}$ as $r\to (2M)_+$. (See Fig. 6.)

\

Replacing (25) in (4) we obtain the metric
\begin{equation}
ds^2_{S_4}|_{I,IV}=- \left(  {{1-{{M}\over{2\rho}}}\over{1+{{M}\over{2\rho}}}} \right)  ^2dt^2+ \left(  1+{{M}\over{2\rho}} \right)  ^4(d\rho^2+\rho^2d\Omega_2).
\end{equation}
$\rho={{M}\over{2}}$ (coordinate singularity corresponding to the horizons) is a fixed point of the \textit{isometry} 
\begin{equation}
\rho\to{{M^2}\over{4\rho}}
\end{equation}
i.e. $ds^2_{S_4}|_{I,IV}(\rho) =ds^2_{S_4}|_{I,IV}({{M^2}\over{4\rho}})$, with $\rho_+\rho_-={{M^2}\over{4}}$, that is $\rho_+\leftrightarrow\rho_-$ under (28).

\

At any hypersurface $t=t_0\in (-\infty,+\infty)$ one obtains the conformally flat metric 
\begin{equation}
-ds^2_{S_4}|_{I,IV;t_0}=(\lambda(\rho))^2(d\rho^2+\rho^2d\Omega_2^2)
\end{equation}
with conformal factor 
\begin{equation}
\lambda(\rho)= \left(  1+{{M}\over{2\rho}} \right)  ^2,
\end{equation}
where 
\begin{equation}
d\rho^2+\rho^2d\Omega_2^2\equiv dl^2_{{\mathbb{E}}^3-\{\vec{0}\}}
\end{equation}
is the Euclidean metric of the punctured 3-space ${\mathbb{R}}^3-\{\vec{0}\}$; it represents a 2-sphere $S^2$ with radius $r=(1+{{M}\over{2\rho}})^2\rho$. Then, topologically,
\begin{equation}
{S_4}|_{I,IV;t_0}\cong \mathbb{R}\times S^2,
\end{equation} 
where at $\rho={{M}\over{2}}$ one has the minimal sphere of radius $2M$. The picture of the wormhole joining the regions $I$ and $IV$ of the Kruskal diagram, is analogous to that corresponding to the case $SaDS_4$ (with $r_h$ replaced by $2M$), except that in the present case the asymptotic space is Euclidean 3-space, while in the $SaDS_4$ case the asymptotic space is Lobachevski (anti De Sitter) 3-space with curvature radius $a$. (See Fig. 7).

\begin{figure}[H]
	\centering
	\includegraphics[width=.7\linewidth]{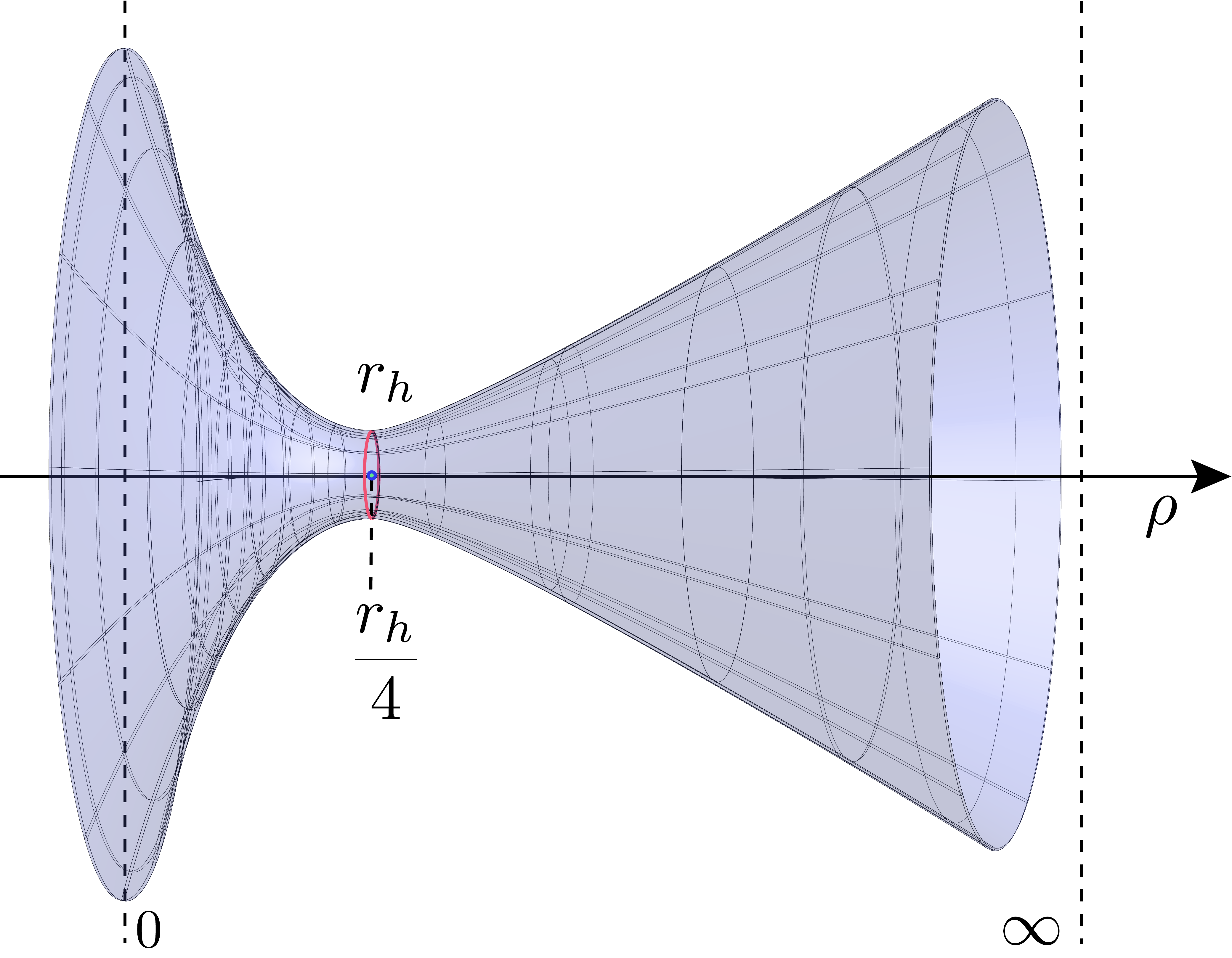}
	\caption{$ SaDS_{4} $ wormhole in isotropic coordinates.}
\end{figure}

\subsection{$SaDS_4$}
The coordinate $\rho\in(0,\infty)$ is now defined as
\begin{equation}
r= \left(  1+{{r_h}\over{4\rho}} \right)  ^2\rho
\end{equation}
with $r_h$ given by (2). As $a\to\infty$, (33) coincides with (25). As before, since $r=\rho+{{r_h}^2\over{16\rho}}+{{r_h}\over{2}}$, for both $\rho\to 0_+$ and $\rho\to+\infty$, $r\to+\infty$. The minimum of $r(\rho)$ is at $\rho={{r_h}\over{4}}$ with value $r=r_h$, and the two roots of (33) for a given $r>r_h$ are 
\begin{equation}
\rho_\pm={{1}\over{2}} \left(  \left(  r-{{r_h}\over{2}} \right)  \pm\sqrt{r(r-r_h)} \right)  .
\end{equation}
The picture of $r(\rho)$ is analogous to that in Fig. 6 with $2M$ replaced by $r_h$ and $M/2$ by $r_h/4$. As in the $S_4$ case, the coordinates $(t,\rho,\theta,\varphi)$ cover only the regions $I$ and $IV$ exterior to the horizons in the Kruskal diagram for $SaDS_4$ (Fig. 8).

\begin{figure}[H]
	\centering
	\includegraphics[width=.9\linewidth]{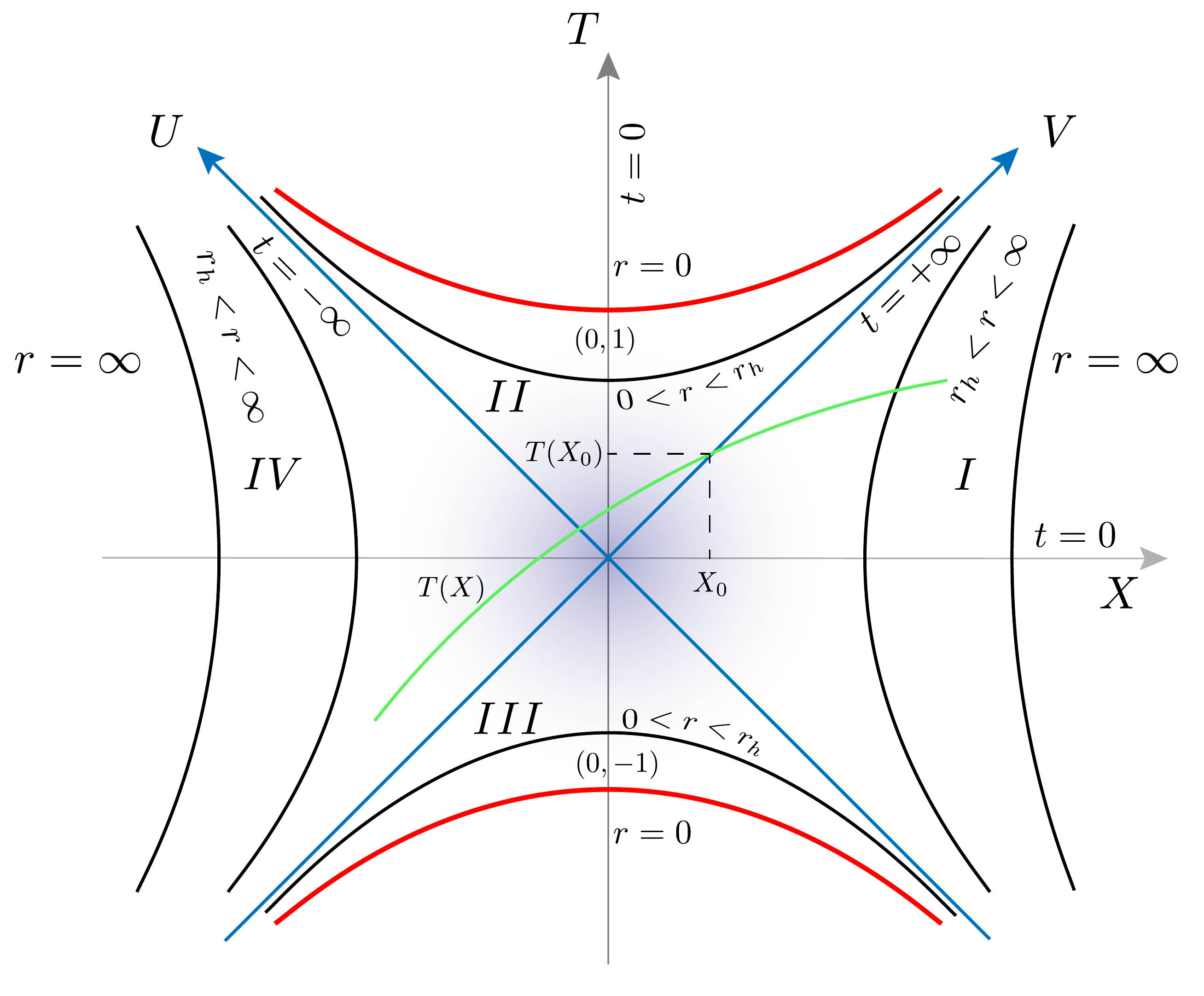}
	\caption{Kruskal-Szekeres diagram of $ SaDS_{4} $.}
\end{figure}

Replacing (33) in (9), the metric results 
\begin{equation}
ds^2_{SaDS_4}|_{I,IV}=Adt^2-{{B}\over{A}}d\rho^2- \left(  1+{{r_h}\over{4\rho}} \right)  ^4\rho^2d\Omega^2_2  
\end{equation}
where
\begin{equation}
A= \left(  {{1-{{r_h}\over{4\rho}}}\over{1+{{r_h}\over{4\rho}}}} \right)  ^2-{{r_h}^3\over{\rho a^2 \left(  1+{{r_h}\over{4\rho}} \right)  ^2}}+ \left(  1+{{r_h}\over{4\rho}} \right)  ^4{{\rho^2}\over{a^2}}
\end{equation}
and
\begin{equation}
B= \left(  1- \left(  {{r_h}\over{4\rho}} \right)  ^2 \right)  ^2.
\end{equation}
As in the $S_4$ case, the transformation 
\begin{equation}
\rho\to{{r_h^2}\over{16\rho}}
\end{equation}
is an \textit{isometry} of (35) i.e.
\begin{equation}
ds^2_{SaDS_4}|_{I,IV} \left(  {{r_h^2}\over{16\rho}} \right) =ds^2_{SaDS_4}|_{I,IV}(\rho),
\end{equation} 
which has $\rho={{r_h}\over{4}}$ as a fixed point. Also, $\rho_+\rho_-={{r_h^2}\over{16}}$, that is $\rho_+\leftrightarrow\rho_-$. 

\

At any fixed Schwarzschild time $t=t_0\in(-\infty,+\infty)$, one obtains an hypersurface with metric 
\begin{equation}
-ds^2_{SaDS_4}|_{I,IV;t_0}={{B}\over{A}}d\rho^2+ \left(  \left(  1+{{r_h}\over{4\rho}} \right)  ^2 \rho \right) ^2 d\Omega_2^2,
\end{equation}
which, at the equator $\theta=\pi/2$ is the 2-surface 
\begin{equation}
-ds^2_{SaDS_4}|_{I,IV;t_0,\pi/4}={{B}\over{A}}d\rho^2+ \left(  \left(  1+{{r_h}\over{4\rho}} \right)  ^2 \rho \right)  ^2 d\varphi^2.
\end{equation} 
(40) ((41)) is a continuum ``sucession" of 2-spheres (1-spheres) with radius $(1+{{r_h}\over{4\rho}})^2\rho$ which goes to infinity when $\rho\to 0_+$ and $\rho\to+\infty$. Asymptotically, i.e. when $r\to+\infty$, as we mentioned in subsection 3.1, (40) ((41)) tends to the Lobachevski 3-space (2-plane) with curvature radius $a$. This is plotted in Fig. 7.
\subsection{$RN_4$}
As in the cases 3.1 and 3.2 we define the coordinate $\rho\in (0,+\infty)$ as 
\begin{equation}
r=(1+r_+/4\rho)^2\rho.
\end{equation}
So, $r\to+\infty$ as $\rho$ when $\rho\to+\infty$, and $r\to+\infty$ as $1/\rho$ when $\rho\to 0_+$. The minimum of $r$ is at $\rho=r_+/4$ with $r(r_+/4)=r_+$. The picture is as that in Fig. 6 with $M/2$ replaced by $r_+/4$ and $2M$ by $r_+$. 

\

The metric (15) becomes 
\begin{multline}
ds^2_{{RN}_4}(t,\rho,\theta,\varphi)=  \left( 1-{{2M}\over{(1+r_+/4\rho)^2\rho}}+{{Q^2}\over{(1+r_+/4\rho)^4\rho^2}} \right) dt^2\\
-{{ \left(  1-(r_+/4\rho)^2 \right) ^2}\over{1-{{2M}\over{(1+r_+/4\rho)^2\rho}}+{{Q^2}\over{(1+r_+/4\rho)^4\rho^2}}}}d\rho^2 -(1+r_+/4\rho)^4\rho^2d\Omega_2^2
\end{multline}
and so at the equator $\theta=\pi/2$ and fixed $t=t_0$,
\begin{equation}
-ds^2_{{RN}_4}|_{t_0,\pi/2}(\rho,\varphi)={{(1-(r_+/4\rho)^2)^2}\over{1-{{2M}\over{(1+r_+/4\rho)^2\rho}}+{{Q^2}\over{(1+r_+/4\rho)^4\rho^2}}}}d\rho^2+((1+r_+/4\rho)^2\rho)^2d\varphi^2.
\end{equation}
As in (38), the transformation
\begin{equation}
\rho\to{{r_+^2}\over{16\rho}}
\end{equation}
is an isometry of the metric i.e. 
\begin{equation}
ds^2_{{RN}_4}(t,{{r_+^2}\over{16\rho}},\theta,\varphi)=ds^2_{{RN}_4}(t,\rho,\theta,\varphi),
\end{equation}  
and obviously also of $-ds^2_{{RN}_4}|_{t_0,\pi/2}$. The coordinates $(t,\rho,\theta,\varphi)$ cover only the regions $I$ and $I^\prime$ exterior to the horizons $r_+$ in the Kruskal diagram for ${RN}_4$ (Fig. 9.a). The asymptotically ($r\to+\infty$) flat hypersurface (surface) described by the metric $-ds^2_{{RN}_4}|_{t_0}(\rho,\theta,\varphi)$ ($-ds^2_{{RN}_4}|_{t_0,\pi/2}(\rho,\varphi)$) consists of a continuum ``succesion" of 2-spheres (1-spheres)with radius $(1+r_+/4\rho)^2\rho$ which $\to+\infty$ as $\rho\to 0_+$ and $\rho\to+\infty$. The picture is analogous to that in Fig. 7 for $SaDS_4$ with $r_h$ replaced by $r_+$ at $\rho=r_+/4$, and asymptotic Euclidean spaces $\mathbb{E}^3$ (planes  $\mathbb{E}^2)$ at $\rho=0,+\infty$.  

\begin{figure}[H]
	\centering
	\includegraphics[width=1\linewidth]{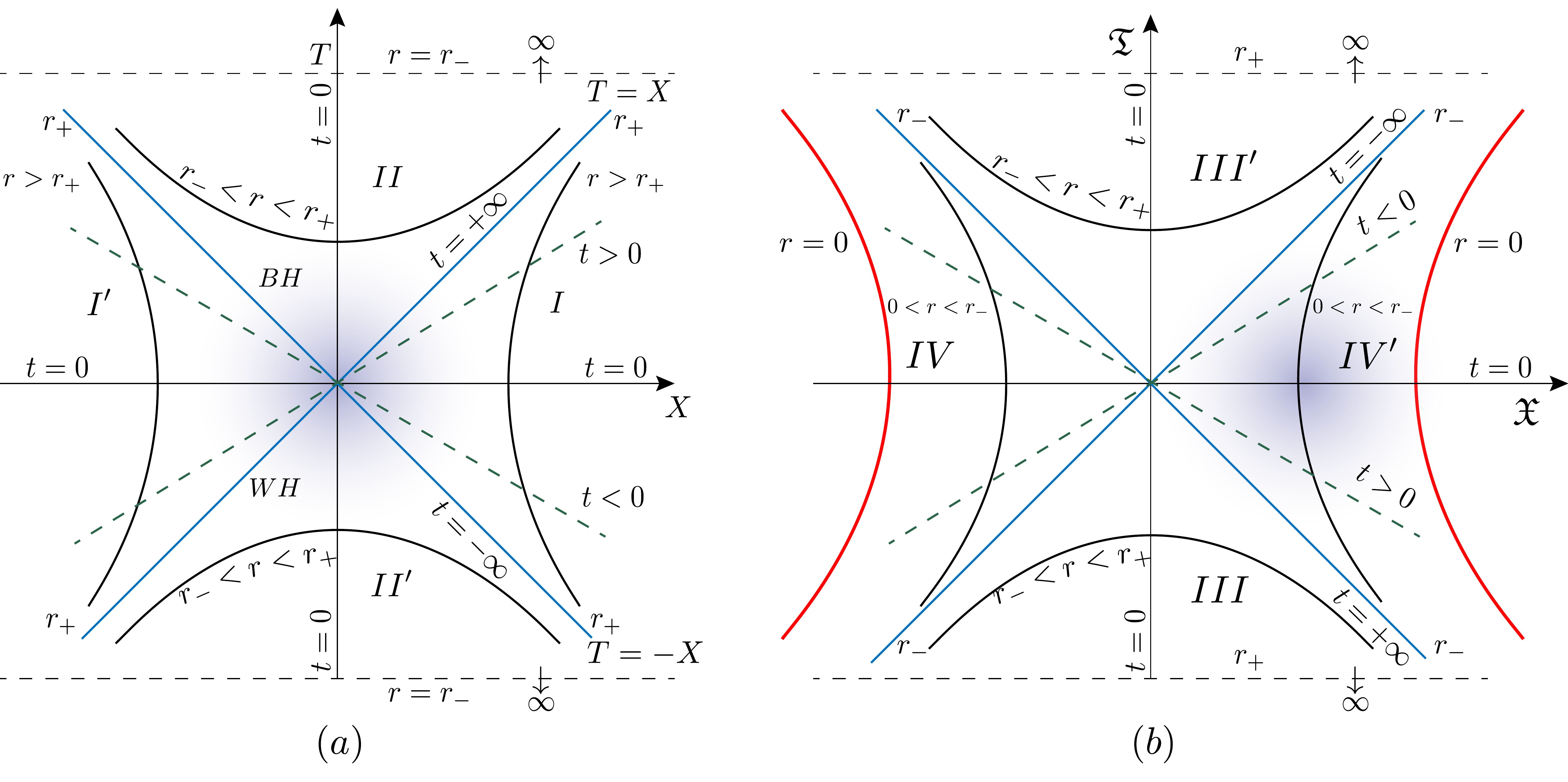}
	\caption{Kruskal-Szekeres diagram of $ RN_{4} $ for the regions: \\ (a) Exterior ($ I,I' $), $ BH(II) $, $ WH(II') $ \\ (b) $ BH(III) $, $ WH(III') $,``singular" $ (IV, IV') $.}
\end{figure}

\subsection{$ERN_4$}
The analysis and Figures for the extreme Reissner-Nördstrom wormhole in isotropic coordinates mimic those for $RN_4$, with the unique replacement $r_+=r_-=M$.
\section{Kruskal-Szekeres coordinates}
\subsection{$SaDS_4$}
The K-S coordinates [1,2] allow the maximal analytic extension of black hole solutions; in particular of the $S_4$, $SaDS_4$ and $RN_4$ metrics. Besides the BH and WH regions, respectively $II$ and $III$ in Figures 2 and 8, respectively for 
$S_4$ and $SaDS_4$, and $II$ and $II^\prime$ in Fig. 9.a for $RN_4$ in the regions $r_-<r<r_+$, they lead to the appearance of the ``other" universe $IV$ for $S_4$ and $SaDS_4$ (Figs. 2 and 8) and $I^\prime$ for $RN_4$ (Fig. 9.a), which, together with the ``Schwarzshild region" $I$, are asymptotically flat for $S_4$ and $RN_4$ and anti De Sitter for $SaDS_4$. The pinching-off and therefore non traversability of the $S_4$ ER bridge is qualitatively discussed in Carroll [6] and quantitatively by Collas and Klein [7], and we refer the reader to them. In this section we discuss in detail the $SAdS_4$, $RN_4$, and $ERN_4$ cases. 

\

The basic strategy to prove the pinching-off of the $SaDS_4$ ER bridge is, like in other cases, to construct the embedding of the bridge in Euclidean space with cylindrical coordinates $(z,r,\varphi)$ for $r<r_h$ and then show that as the time-radial K-S coordinates $(T,X)\to(1,0)$ in the BH region (or to $(-1,0)$ in the WH region) -and therefore to the singularity $r=0$ (see Fig. 8)- for the embedding function $z=z(r)$ one has $({{dz}\over{dr}})^2\to+\infty$ i.e. ${{dz}\over{dr}}\to\pm\infty$. This implies that the revolution surface (or hypersurface if the metric is not restricted to the equator) pinches-off and therefore makes the wormhole non traversable.

\

Let 
\begin{equation}
f(r)=1-{{2M}\over{r}}+{{r^2}\over{a^2}};
\end{equation}  
then
\begin{equation}
f^\prime(r_h)={{df}\over{dr}}|_{r=r_h}=2 \left(  {{M}\over{r_h^2}}+{{r_h}\over{a^2}} \right)  .
\end{equation}
Defining the coordinates 
\begin{equation}
U=-sign(f)e^{f^\prime(r_h)(r^*(r)-t)/2)}, \ V=e^{f^\prime(r_h)(r^*(r)+t)/2)}
\end{equation}
with $U<0$ for $r>r_h$ ($I$), $U>0$ for $r<r_h$ ($II$), and $V>0$ for all $r$, where $r^*(r)$ is the tortoise Regge-Wheeler radial coordinate given by 
\begin{multline}
r^*(r)=\int_0^r{{dr^\prime}\over{f(r^\prime)}} ={{a^2}\over{3r_h^2+a^2}}(r_hln|1-{{r}\over{r_h}}|-{{r_h}\over{2}}ln(1+{{r(r+r_h)}\over{r_h^2+a^2}})\\
+{{3r_h^2+2a^2}\over{\sqrt{3r_h^2+4a^2}}}arctg({{r\sqrt{3r_h^2+4a^2}}\over{2(r_h^2+a^2)+rr_h}}))           
\end{multline}
[14]. From (49), $r^*(r)={{1}\over{f^\prime(r_h)}}ln(-sgn(f)UV)=r^*(UV)=r^*(U,V)$, with the symmetry 
\begin{equation}
r^*(-U,-V)=r^*(U,V).
\end{equation} 
For both $r>r_h$ and $r<r_h$, $r$ is unambiguously determined from $r^*$; then $r(U,V)=r(-U,-V)$. Then $(U,V)\to(-U,-V)$ is a symmetry of the $SaDS_4$ metric
\begin{equation}
ds^2_{SaDS_4}=-4{{f(r(U,V))}\over{(f^\prime(r_h))^2}}{{dUdV}\over{UV}}-r^2(U,V)d\Omega^2_2,
\end{equation}
which can be extended to the regions $IV$ ($U>0$, $V=-e^{f^\prime(r_h)(r^*(r)+t)/2)}<0$) and $III$ ($U<0)$, $V=-e^{f^\prime(r_h)(r^*(r)+t)/2)}<0$). To pass from the two null coordinates $(U,V)$ and two spacelike coordinates $(\theta,\varphi)$ to one timelike coordinate ($T$) and three spatial coordinates $(X,\theta,\varphi)$, we define 
\begin{equation}
T={{V+U}\over{2}}, \ X={{V-U}\over{2}}
\end{equation}
(both $T,X\in(-\infty,+\infty)$ and the metric becomes 
\begin{equation}
ds^2_{SaDS_4}(T,X,\theta,\varphi)=-4{{f(r(T,X))}\over{(f^\prime(r_h))^2}}{{dT^2-dX^2}\over{T^2-X^2}}-r^2(T,X)d\Omega^2_2(\theta,\varphi).
\end{equation}
For a spatial section $T=T(X)$, (54) becomes 
\begin{multline}
ds^2_{SaDS_4}(T(X),X,\theta,\varphi)=\\
-4{{f(r(T(X),X))}\over{(f^\prime(r_h))^2}}{{dX^2}\over{T(X)^2-X^2}}((T^\prime(X))^2-1)-r^2(T(X),X)d\Omega^2_2(\theta,\varphi)
\end{multline}
with $T^\prime(X)={{dT(X)}\over{dX}}$. Again by spherical symmetry at each spacetime point we can choose $\theta=\pi/2$ and then 
\begin{multline}
-ds^2_{SaDS_4}(T(X),X,\pi/2,\varphi)=\\
4{{f(r(T(X),X))}\over{(f^\prime(r_h))^2}}{{dX^2}\over{T(X)^2-X^2}}((T^\prime(X))^2-1)+r^2(T(X),X)d\varphi^2.
\end{multline}
It is clear that this expression is the metric of a 2-dimensional surface. To proceed to its embedding in $\mathbb{E}^3$, we define the vector 
\begin{equation}
\vec{x}=\vec{x}(X,\varphi)=(F(X)cos\varphi,F(X)sin\varphi,z(X))
\end{equation}
with squared length 
\begin{equation}
dl_{\vec{x}}^2=((F^\prime(X))^2+(z^\prime(X))^2)dX^2+F(X)^2d\varphi^2.
\end{equation}
Identifying (56) and (58),
\begin{equation}
(F^\prime(X))^2+(z^\prime(X))^2=4{{f(r(T(X),X))}\over{(f^\prime(r_h))^2}}{{(T^\prime(X))^2-1}\over{T(X)^2-X^2}},
\end{equation}
\begin{equation}
(F(X))^2=(r(T(X),X))^2
\end{equation}
i.e. $F(X)=r(X)$. Using
\begin{equation}
(T(X))^2-X^2=\mp e^{f^\prime(r_h)r^*(r)}
\end{equation}
where the -(+) sign corresponds to $r>r_h$ ($r<r_h$), (59) becomes 
\begin{equation}
(r^\prime(X))^2+(z^\prime(X))^2=\mp 4{{f(X)e^{-f^\prime(r_h)r^*(r)}}\over{(f^\prime(r_h))^2}}(T^\prime(X))^2-1).
\end{equation}
For a constant spatial section $T(X)=T_0=const.$,
\begin{equation}
(r^\prime(X))^2+(z^\prime(X))^2=\pm 4{{f(X)e^{-f^\prime(r_h)r^*(r)}}\over{(f^\prime(r_h))^2}}.
\end{equation}
Differentiating both sides of (61) at $T=T_0$, ${{dX}\over{dr}}|_{T=T_0}=\pm{{1}\over{2}}{{f^\prime(r_h)e^{f^\prime(r_h)r^*(r)}}\over{X}}{{dr^*}\over{dr}}$, and using ${{dz}\over{dX}}={{dz}\over{dr}}{{dr}\over{dX}}$ ie. ${{dz}\over{dr}}={{dz}\over{dX}}{{dX}\over{dr}}$, one obtains, for the case $r<r_h$, 
\begin{equation}
({{dz}\over{dr}})^2|_{r=\epsilon} =-{{e^{f^\prime(r_h)r^*(r)}}\over{f(X)X^2}}-1.
\end{equation}
From Figures 8 and 10 in [14], for $r<r_h$, both $f(r)$ and $r^*(r)$ are negative, with $f(r)$ decreasing with decreasing $r$ and with $f(r)\to -\infty$ as $r\to 0_+$ and $f(r)\to 0_-$ as $r\to {r_h}_-$, while $r^*(r)$ decreases with increasing $r$, with $r^*(r)=0$ and $r^*(r)\to -\infty$ as $r\to {r_h}_-$. The r.h.s. of (64) is positive or null if $e^{-f^\prime(r_h)r^*(r)}\leq 1+1/|f(r)|$. Both the left and right hand sides of this inequality $\to 1_+$ when $r\to 0_+$ and $\to +\infty$ when $r\to {r_h}_-$. As $r\to 0_+$, the slope of the l.h.s. $\to 0_+$ while that of the r.h.s. $\to 1_+$, and as $r\to {r_h}_-$ the slope of the l.h.s. $\to +\infty$ exponentially while that of the r.h.s. $\to +\infty$ as $1/\delta^2$ for $\delta\to 0$. Then the two curves intersect each other and the sign of the inequality changes. This means that there exists a unique $\bar{r}\in(0,r_h)$ such that the inequality holds for $0\leq r\leq\bar{r}$.

\

Given $T=T_0$, a choice of $X$ fixes $r=r(T_0,X)$ (see Fig. 8). Let's choice $T_0=1$; then as $X\to 0_+$, $r\to 0_+$. From (50), $r^*(0)=0$ and for small $r=\epsilon$, $r^*(\epsilon)\cong{{\epsilon}\over{f(\epsilon)}}\cong-{{\epsilon^2}\over{2M}}$. On the other hand, from (61), $1=X^2+e^{f^\prime(r_h)r^*(r)}$ and so $X^2=1-e^{f^\prime(r_h)r^*(r)}\cong 1-e^{-f^\prime(r_h){{\epsilon^2}\over{2M}}}\cong f^\prime(r_h){{\epsilon^2}\over{2M}}.$ Then, 
\begin{equation}
({{dz}\over{dr}})^2|_{r=\epsilon}\cong {{1}\over{f^\prime(r_h)\epsilon}}\to+\infty
\end{equation}
as $\epsilon\to 0_+$. So, 
\begin{equation}
{{dz}\over{dr}}\to\pm\infty.
\end{equation}
Then the embedding curve $z(r)$ at $T=X=0$ and therefore at $r=0$ has the behavior shown in Fig. 10. (The fact that $r>0$ excludes the existence of an inflection point.) It is clear that the revolution surface generated by it as $\varphi:0\to 2\pi$ pinches-off and makes the $SaDS_4$ wormhole non traversable. 

\begin{figure}[H]
	\centering
	\includegraphics[width=.5\linewidth]{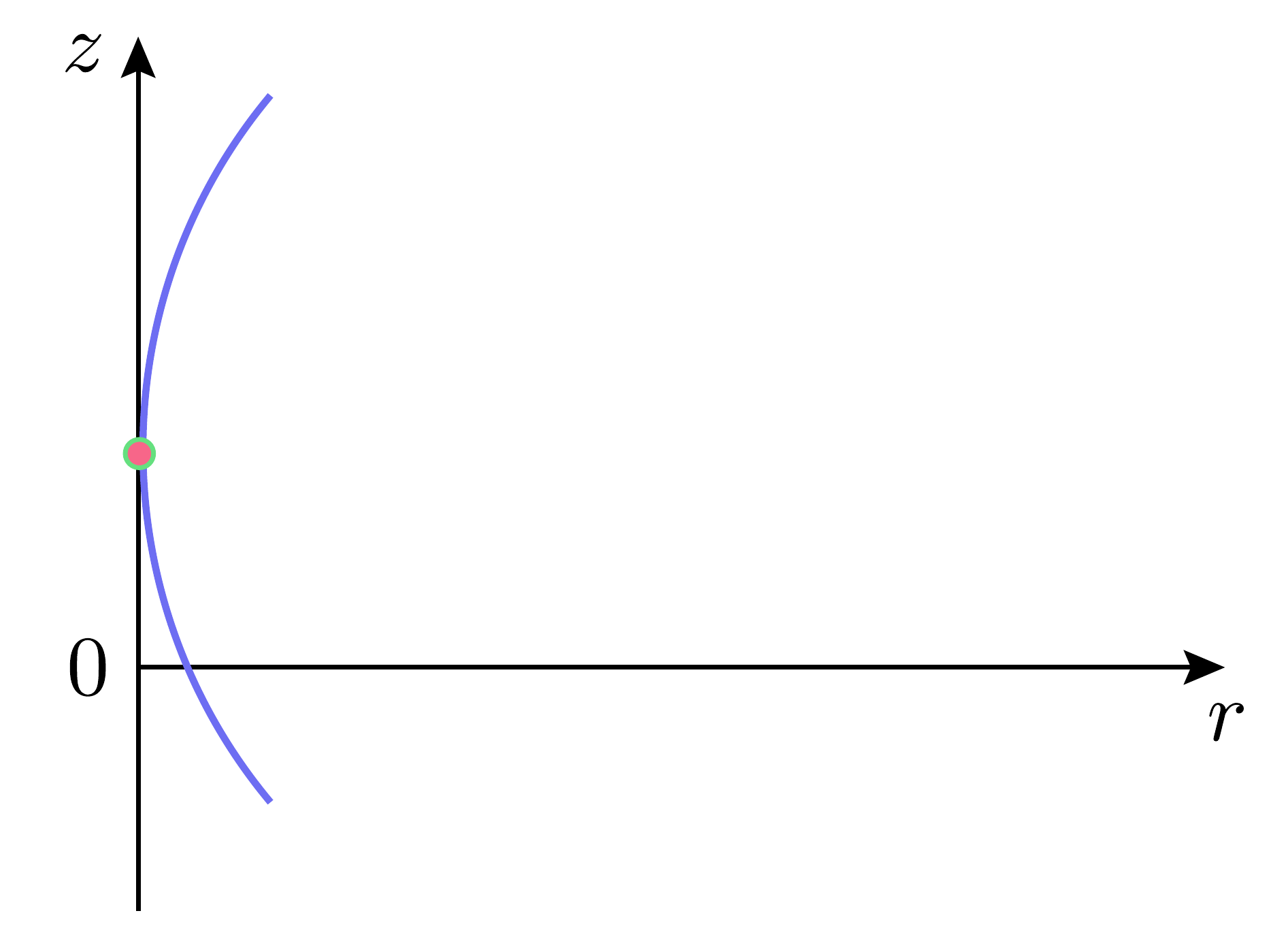}
	\caption{Embedding curve $ z(r) $ for $ SaDS_{4} $ near $ r=0 $.}
\end{figure}

\subsection{$RN_4$}
In Kruskal-Szekeres coordinates $(T,X,\theta,\varphi)$ the $RN_4$ metric in the regions $r_+\leq r$ (exterior $I$ and $I^\prime$) and $r_-<r\leq r_+$ (BH $II$ and WH $II^\prime$) is given by 
\begin{equation}
ds^2_{RN_4}={{2M}\over{\kappa_+^2r^2}}e^{-2\kappa_+r}({{r-r_-}\over{2M}})^{(r_-/r_+)^2}(r-r_-)(dT^2-dX^2)-r^2d\Omega_2^2
\end{equation}
where $r_\pm$ are given in (17), 
\begin{equation}
\kappa_{\pm}=\pm{{r_+-r_-}\over{2r_\pm^2}}
\end{equation}
are the corresponding surface gravities, and $r=r(T,X)$ is implicitly given by
\begin{equation}
T^2-X^2=\mp{{e^{2\kappa_+r}|r-r_+|(2M)^{(r_-/r_+)^2-1}}\over {(r-r_-)^{(r_-/r_+)^2}}}
\end{equation}
with - sign for $r_+\leq r$ and + sign for $r_-<r\leq r_+$. The K-S diagram is shown in Fig. 9a. 

\

Let us first study the wormhole embedding in the regions $r_-<r\leq r_+$. If $T=T(X)$ is a spacelike section, the metric at a fixed $T(X)=T_0$ and $\theta=\pi/2$ is 
\begin{equation}
ds^2_{RN_4}|_{T_0,\pi/2}={{2M}\over{\kappa_+^2r^2}}e^{-2\kappa_+r}({{r-r_-}\over{2M}})^{(r_-/r_+)^2}(r-r_-)dX^2-r^2d\varphi^2.
\end{equation}
Defining the vector in $\mathbb{E}^3$
\begin{equation}
\vec{x}=\vec{x}(X,\varphi)=(G(X)cos\varphi,G(X)sin\varphi,z(X))
\end{equation}  
with 
\begin{equation}
dl_{\vec{x}}^2=((G^\prime(X))^2+(z^\prime(X))^2)dX^2+(G(X))^2d\varphi^2
\end{equation}
and identifying $-ds^2_{RN_4}|_{T_0,\pi/2}$ with $dl_{\vec{x}}^2$, one obtains $G(X)=r(X)$ and 
\begin{equation}
({{dz}\over{dr}})^2=-{{e^{-2\kappa_+r}}\over{\kappa_+^2r^2}}{{(r-r_-)^{(r_-/r_+)^2+1}}\over{(2M)^{(r_-/r_+)^2-1}}}({{dX}\over{dr}})^2-1
\end{equation}
where we used ${{dz}\over{dX}}={{dz}\over{dr}}{{dr}\over{dX}}$. To obtain $({{dX}\over{dr}})^2$ we use (69) with $T=T_0=0$:
\begin{equation}
({{dX}\over{dr}})^2={{(2M)^{2((r_-/r_+)^2-1)}}\over{4X^2}}{{d}\over{dr}}({{e^{2\kappa_+r}(r_+-r)}\over{(r-r_-)^{(r_-/r_+)^2}}})^2,
\end{equation}
\begin{equation}
X^2=-{{e^{2\kappa_+r}(r_+-r)(2M)^{(r_-/r_+)^2-1}}\over {(r-r_-)^{(r_-/r_+)^2}}},
\end{equation}
which leads to
\begin{equation}
({{dz}\over{dr}})^2={{1}\over{4\kappa_+^2r^2}}({{r-r_-}\over{r_+-r}})(2\kappa_+(r_+-r)-(r_-/r_+)^2({{r_+-r}\over{r-r_-}})-1)^2-1.
\end{equation}
As $r\to(r_-)_+$, $r-r_-=\epsilon$ with $|\epsilon|<<1$, $\epsilon\to 0_+$; so the r.h.s. of (76) $\sim\epsilon\times{{1}\over{\epsilon^2}}={{1}\over{\epsilon}}\to+\infty$ i.e. 
\begin{equation}
({{dz}\over{dr}})^2\to+\infty \ as \ r\to (r_-)_+,
\end{equation} 
which implies
\begin{equation}
{{dz}\over{dr}}\to\pm\infty \ as \ r\to(r_-)_+.
\end{equation} 
The form of the embedding curve is shown in Fig. 11, giving at this point a wormhole radius $r_-$. 

\begin{figure}[H]
	\centering
	\includegraphics[width=.7\linewidth]{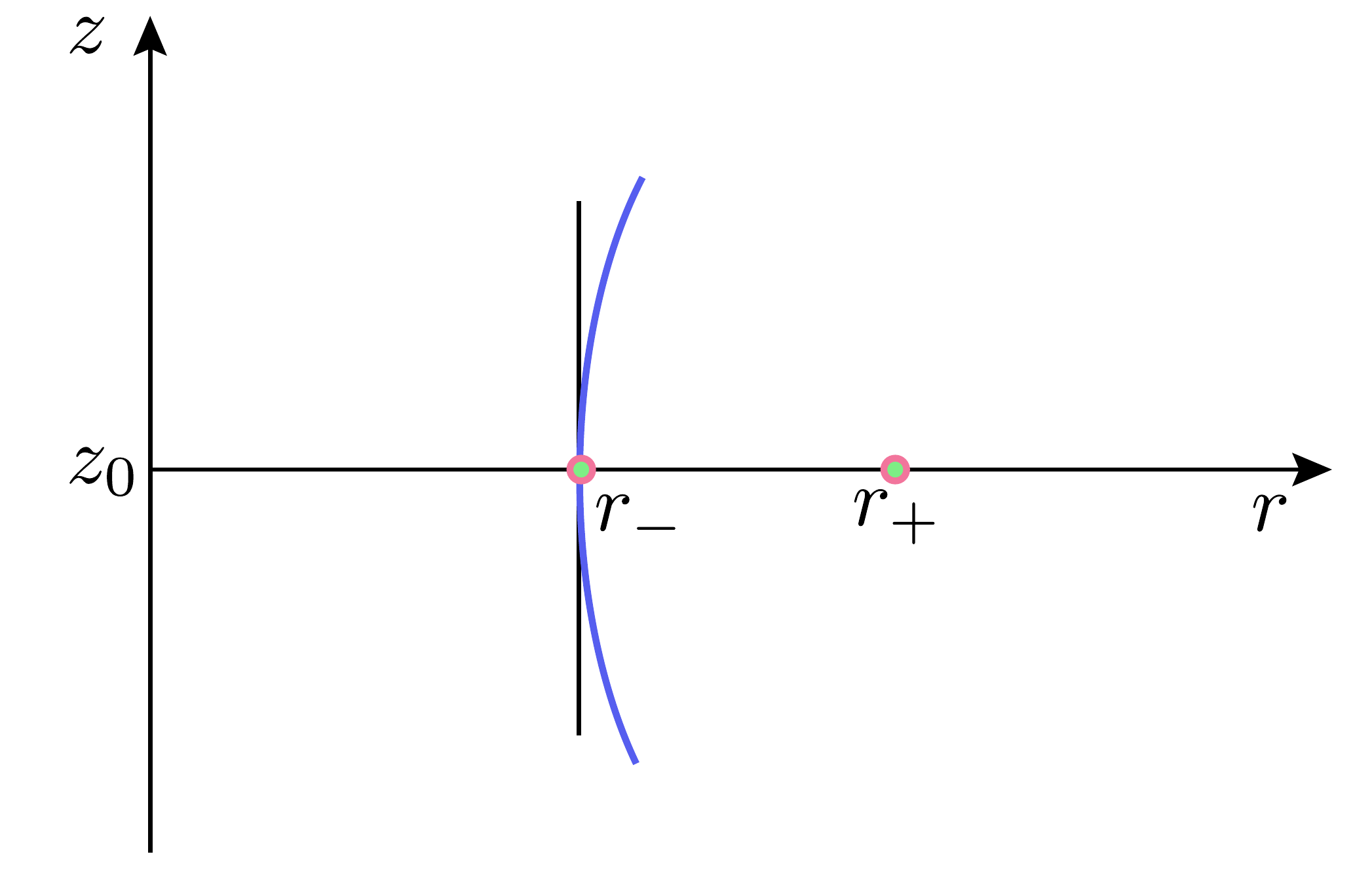}
	\caption{Embedding curve $ z(r) $ at $ r=r_{-} $ for $ RN_{4} $.}
\end{figure}

For the embedding in the regions $IV$ and $IV^\prime$ with boundaries at $r=r_-$ and at $r=0$ (timelike singularity hypersurface) i.e. for $0<r\leq r_-$, we use the K-S coordinates $(\mathfrak{T},\mathfrak{X},\theta,\varphi)$ which cover the regions $IV$, $IV^\prime$, $III$ and $III^\prime$ (Fig. 9b) with metric 
\begin{equation}
d\tilde{s}^2_{RN_4}={{2M}\over{\kappa_-^2r^2}}e^{-2\kappa_-r}({{r_+-r}\over{2M}})^{(r_+/r_-)^2}(r_+-r)(d\mathfrak{T}^2-d\mathfrak{X}^2)-r^2d\Omega_2^2,
\end{equation}
where $r=r(\mathfrak{T},\mathfrak{X})$ is implicitly given by 
\begin{equation}
\mathfrak{T}^2-\mathfrak{X}^2=-{{e^{2\kappa_-r}|r-r_-|(2M)^{(r_+/r_-)^2-1}}\over {(r_+-r)^{(r_+/r_-)^2}}}.
\end{equation} 
(The region $II$ in Fig. 9a coincides with the region $III$ in Fig. 9b, and the same holds between the regions $II^\prime$ in Fig. 9a and $III^\prime$ in Fig. 9b; the transformations $(T_{II}(t,r),X_{II}(t,r))\to (\mathfrak{T}_{III}(t,r),\mathfrak{X}_{III}(t,r))$ and $(T_{II^\prime}(t,r),X_{II^\prime}(t,r))\to (\mathfrak{T}_{III^\prime}(t,r),\mathfrak{X}_{III^\prime}(t,r))$ are diffeomorphisms.)

\

Again, if $\mathfrak{T}=\mathfrak{T}(\mathfrak{X})$ is a spacelike section, the metric at a fixed $\mathfrak{T}=\mathfrak{T}_0$ and at the equator $\theta=\pi/2$ becomes
\begin{equation}
d\tilde{s}^2_{RN_4}|_{\mathfrak{T}_0,\pi/2}=-{{2M}\over{\kappa_-^2r^2}}({{r_+-r}\over{2M}})^{(r_+/r_-)^2}(r_+-r)d\mathfrak{X}^2-r^2d\varphi^2.
\end{equation}
For its embedding in $\mathbb{E}^3$ we use $dl_{\vec{x}}^2$ given by (72), and eq. (80) at $\mathfrak{T}_0=0$; a straightforward calculation similar to the one for the case $r_-<r\leq r_+$ leads to
\begin{equation}
({{dz}\over{dr}})^2={{1}\over{4\kappa_-^2r^2}}({{r_+-r}\over{r_--r}})((2\kappa_-+{{(r_+/r_-)^2}\over{r_+-r}})(r_--r)-1)^2-1.
\end{equation}
Near the singularity i.e. for $r=\epsilon$, $|\epsilon|<<1$,
\begin{equation}
({{dz}\over{dr}})^2\cong {{r_+/r_-}\over{4\kappa_-^2\epsilon^2}}(2\kappa_-r_-+r_+/r_--1)^2-1={{r_+r_-}\over{4\epsilon^2}}-1\to+\infty \ as \ \epsilon\to 0_+
\end{equation}
and therefore
\begin{equation}
{{dz}\over{dr}}\to\pm\infty \ as \ \epsilon\to 0_+.
\end{equation}
So, the form of the embedding curve $z(r)$ is like that in the $SaDS_4$ case (Fig. 10), meaning that the wormhole pinches-off at the singularities becoming non traversable.   
\subsection{$ERN_4$}
There is only one singularity at $r=0$ (Fig. 5) and the result is the same as for $RN_4$. In (83), $r_=r_+=M$, and so 
\begin{equation}
({{dz}\over{dr}})^2={{r_+^2}\over{4\epsilon^2}}-1\to +\infty \ as \ \epsilon\to 0_+,
\end{equation}
and therefore
\begin{equation}
{{dz}\over{dr}}\to\pm\infty \ as \ \epsilon\to 0_+.
\end{equation}

\section{Final remarks}
Though we have reviewed the kinematical reason for the non traversability of the wormholes associated to the black holes $S_4$, $SaDS_4$, and $RN_4$, intensive work is being doing at present on eternal traversable black holes [15,16], with appropiate quantum conditions on the energy-momentum tensor; in particular, the violation of the average null energy condition. It is well known that through a Casimir-like effect, quantum fields can produce states with negative energy at a given spacetime point, stabilizing an otherwise non traversable wormhole. According to the present authors, the most interesting problem is to see if some of these efforts towards the construction of traversable wormholes can lead to the existence of closed causal curves and therefore to time travel.

\section*{Acknowledgment}
M.S. thanks the Alexander von Humboldt Foundation for a Research Fellowship, and to the Carl Friedrich von Siemens Foundation for partial support.

\section*{References}
 
 \
 
 [1]. Kruskal, M.D. \textit{Maximal extension of Schwarzschild metric}, Phys. Rev. \textbf{119}, 1743-1745 (1960).
 
 \  
   
 [2]. Szekeres, G. \textit{On the singularities of a Riemannian manifold}, Publ. Math. Debrecen \textbf{7}, 285-301 (1960). 
  
  \ 
   
 [3]. Misner, C.W. and Wheeler, J.A. \textit{Classical Physics as Geometry}, Ann. of Phys. \textbf{2}, 525-603 (1957).
   
  \

 [4]. Einstein, A. and Rosen, N. \textit{The particle problem in the general theory of relativity}, Phys. Rev. \textbf{48}, 73-77 (1935).  
   
  \

 [5]. Fuller, R.W. and Wheeler, J.A. \textit{Causality and Multiply Connected Space-Time}, Phys. Rev. \textbf{128}, 919-929 (1962).

   \

 [6]. Carroll, S. ``Spacetime and Geometry. An Introduction to General Relativity", Addison Wesley, San Francisco (2004).

   \
   
 [7]. Collas, P. and Klein, D. \textit{Embeddings and time evolution of the Schwarzschild wormhole}, Am. J. Phys. \textbf{80}, 203-210 (2012).

   \

 [8]. Morris, M.S. and Thorne, K.S. \textit{Wormholes in spacetime and their use for interstellar travel: A tool for teaching General Relativity}, Am. J. Phys. \textbf{56}, 395-412 (1988).

   \

 [9]. Visser, M. ``Lorentzian Wormholes: From Einstein to Hawking", American Institute of Physics, New York (1995).

  \

 [10]. Lobo, F.S.N. ``From the Flamm-Einstein-Rosen bridge to the modern renaissence of traversable wormholes", The Fourteenth Marcel Grossmann Meeting, University of Rome ``La Sapienza", World Scientific, 409-427 (2017).

   \

 [11]. Witten, E. \textit{Light Rays, Singularities, and All That}, arXiv: hep-th/1901.03928v1 (2019).

  \

 [12]. Flamm, L. \textit{Beiträge zur Einsteinschen Gravitationtheorie}, Physikalische Zeitschrift \textbf{XVII}, 448-454 (1916); reprinted: \textit{Contributions to Einstein's theory of gravitation}, Gen. Relat. Gravit. \textbf{47}:72 (2015).

   \

 [13]. Townsend, P.K. ``Black Holes", Lecture notes, DAMTP, Cambridge (1997); arXiv: gr-qc/9707012v1.

  \

 [14]. Socolovsky, M. \textit{Schwarzschild Black Hole in Anti-De Sitter Space}, Adv. App. Clifford Algebras \textbf{28}:18 (2018).

  \

 [15]. Gao,P., Jafferis,D.L., and Wall, A.C. \textit{Traversable Wormholes via a Double Trace Deformation}, arXiv: hep-th/1608.05687v2 (2017).

   \

 [16]. Maldacena, J. and Qi, X.L. \textit{Eternal traversable wormhole}, arXiv: hep-th/1804.00491v3 (2018).

\end{document}